\pdfoutput=1
\documentclass[
reprint,
aps,
prd,
superscriptaddress
]{revtex4-2}

\usepackage{graphicx}% Include figure files
\usepackage{dcolumn}% Align table columns on decimal point
\usepackage{bm}% bold math

\newcommand{\beq}{\begin{equation}}
\newcommand{\beqa}{\begin{eqnarray}}
\newcommand{\eeq}{\end{equation}}
\newcommand{\eeqa}{\end{eqnarray}}

\newcommand{\bfx}{\mathbf{x}}

\newcommand{\simgt}{\lower.5ex\hbox{$\; \buildrel > \over \sim \;$}}
\newcommand{\simlt}{\lower.5ex\hbox{$\; \buildrel < \over \sim \;$}}

\newcommand{\bd}[1]{\mbox{\boldmath $#1$}}

\usepackage{xcolor}

\usepackage{amsmath,amssymb,natbib,latexsym,times}

\begin{document}

%\preprint{APS/123-QED}

\title{Recovering Missing Degrees of Freedom \\
in Simulation-Trained Field-Level Generative Models}% Force line breaks with \\
%\title{Variance Shrinkage Reveals Missing Degrees of Freedom in Simulation-Based Generative Inference}
%\title{Missing Degrees of Freedom in Simulation-Trained Generative Models for Field-Level Inference}
%\thanks{hoge}%

\author{Masato Shirasaki}
\email{masato.shirasaki@nao.ac.jp}
\affiliation{National Astronomical Observatory of Japan (NAOJ), National Institutes of Natural Sciences, Tokyo 181-8588, Japan}
\affiliation{The Institute of Statistical Mathematics, Tokyo 190-8562, Japan}
\affiliation{RIKEN Center for Advanced Intelligence Project, Tokyo 103-0027, Japan}

\author{Shiro Ikeda}%
%\email{Second.Author@institution.edu}
\affiliation{The Institute of Statistical Mathematics, Tokyo 190-8562, Japan}

\date{\today}% It is always \today, today,
             %  but any date may be explicitly specified

\begin{abstract}
Simulation-trained generative models are emerging as a powerful framework for field-level inference in cosmology, particularly in strongly nonlinear regimes where analytical approaches become intractable. However, the finite size of training datasets often leads to variance shrinkage, indicating that part of the realization-level variability of the underlying physical process is not captured by the learned generative dynamics. In this paper, we interpret variance shrinkage as a diagnostic of missing effective degrees of freedom and introduce physics-informed latent modifications to the generative dynamics of a continuous normalizing flow. 
We show that these latent corrections recover a substantial fraction of the missing variability without requiring additional training simulations. Using weak-lensing field-level inference as a representative cosmological application, we further demonstrate through posterior analysis that the learned latent directions are actively exploited during inference and behave as effective nuisance parameters that absorb part of the mismatch between the training and target distributions.
As a result, the model exhibits improved robustness to out-of-distribution shifts in field-level inference. 
Our work establishes variance shrinkage as a practical diagnostic for identifying missing degrees of freedom in simulation-trained generative models and provides a framework for recovering them through physics-informed modifications of generative dynamics.
%Our results suggest a general strategy for simulation-based scientific machine learning: rather than relying solely on larger training datasets, missing physical variability can be recovered through physics-informed latent degrees of freedom embedded in the generative dynamics.
\end{abstract}

%\keywords{Suggested keywords}%Use showkeys class option if keyword
                              %display desired
\maketitle

%\tableofcontents

\section{Introduction}

Field-level inference provides a lossless framework for extracting scientific information from observational data, but it requires highly accurate generative models of the underlying physical fields. In modern cosmology, field-level generative models are often constructed using analytical approximations that remain valid on scales larger than $\sim 10\,\mathrm{Mpc}$ \cite{Jasche:2012kq, Schmidt:2020viy, Leclercq2025}. Such approaches become inaccurate in the strongly nonlinear regime, where simulation-trained generative models have emerged as a promising alternative. Recent advances in deep generative modeling have made it possible to emulate highly nonlinear cosmological fields and to perform field-level inference directly from simulation-based priors.

A fundamental challenge of simulation-trained generative models is that the available training sets are inevitably finite. While modern generative models often reproduce the average properties of the training simulations, they frequently underestimate realization-to-realization variability. This phenomenon, which we refer to as variance shrinkage, has recently been identified as an important limitation of simulation-based generative modeling \cite{Shirasaki:2023nnk, Armijo:2026ucb}. More generally, matching low-order summary statistics or posterior means does not guarantee an accurate reconstruction of uncertainty structures in field-level inference \cite{Doeser:2026gnj}. Simply increasing the number of training simulations is often impractical because high-fidelity cosmological simulations remain computationally expensive. Recent reviews of simulation-based inference have therefore identified limited simulation budgets as one of the major obstacles for precision cosmological analyses \cite{Thiele2026}.

In this work, we take a different approach. Rather than viewing variance shrinkage solely as a limitation of finite training data, we interpret it as evidence that the trained generative model is missing effective degrees of freedom required to describe realization-level variability. This interpretation suggests a practical strategy: identify the missing variability through discrepancies between simulations and model predictions, and recover it through physically motivated modifications of the generative dynamics.

To demonstrate this idea, we introduce physics-informed latent corrections into a continuous normalizing flow trained on weak-lensing convergence maps. The latent corrections are motivated by super-survey density fluctuations and are designed to restore variability that is absent in the baseline model. We show that these latent degrees of freedom substantially improve the reproduction of realization-to-realization scatter without degrading the average statistical properties of the generated fields. Through field-level inference, we further demonstrate that the latent variables are actively utilized when confronting out-of-distribution data and behave as effective nuisance directions that absorb part of the mismatch between the training and target distributions.

Our results suggest a broader framework in which physics-informed latent dynamics complement data-driven generative models. In this picture, the generative model learns the complex average behavior encoded in finite simulation sets, while latent corrections restore physically meaningful variability that is difficult to learn from data alone. We argue that this provides a promising pathway toward robust simulation-based field-level inference in cosmology and other scientific applications.

\section{Data and Methods}

\subsection{Weak-lensing simulations}

As a representative cosmological application, we consider field-level modeling of weak gravitational lensing maps. Our goal is to generate convergence fields $\kappa$ for four source-redshift bins centered at $z_{\rm s}=0.5$, $0.75$, $1.0$, and $1.3$, matching a typical configuration of current-generation imaging surveys \cite{Rau:2022wrq}. Brief introductions to weak-lensing convergence fields and the summary statistics adopted in this work are provided in Appendix~\ref{apdx:lensing}.

To train our generative model, we use the publicly available weak-lensing simulations of Ref.~\cite{Osato:2020sxo}, known as the $\kappa$TNG suite. The suite contains 10000 realizations of $5\times5\,\mathrm{deg}^2$ convergence maps for 40 source redshifts up to $z_s=2.6$. Each realization is generated from a pair of parent simulations: a gravity-only run and a full-physics TNG300-1 run that includes baryonic processes such as star formation, gas cooling, chemical enrichment, and black-hole feedback \cite{Pillepich:2017fcc}. Since the two runs share identical initial phases, baryonic effects can be quantified on a realization-by-realization basis.

We train the generative model to produce convergence maps in a $1.25\times1.25\,\mathrm{deg}^2$ field of view for four source-redshift bins centered at $z_s=0.506$, $0.749$, $1.034$, and $1.302$. This tomographic configuration is motivated by the Subaru Hyper Suprime-Cam survey \cite{Rau:2022wrq} and forthcoming Stage-IV weak-lensing surveys such as Euclid and LSST. For training, we use the gravity-only simulations and degrade the original $1024^2$ maps to $256^2$ pixels. To reduce the impact of Fourier modes larger than the target field of view and stabilize the training procedure, we additionally apply a large-scale filtering operation before training. Details of the data preprocessing and filtering procedure are described in Appendix~\ref{apdx:preprocess}.

We train the fiducial model using 10,000 realizations of gravity-only weak-lensing simulations from $\kappa$TNG \cite{Osato:2020sxo}. While the trained flow accurately reproduces the mean statistical properties of the simulations, it exhibits variance shrinkage, namely an underestimation of realization-to-realization variability. This motivates the latent corrections introduced below.

\begin{figure*}[!t]
\includegraphics[clip, width=1.8\columnwidth]{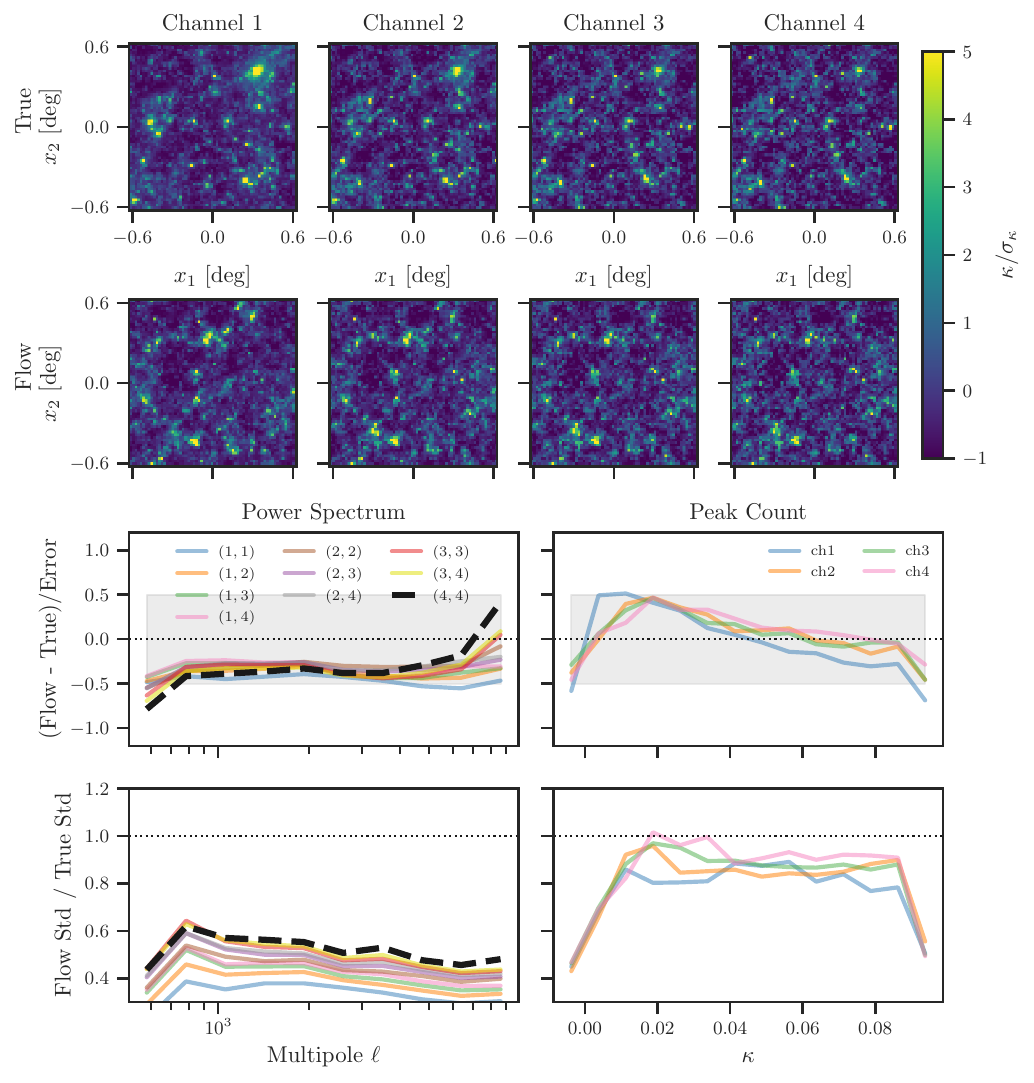}
\caption{\label{fig:baseline_flow} 
Validation of the baseline flow model. The first row shows a representative realization 
from the $\kappa$TNG simulations, while the second row shows a sample generated by the baseline flow model. Field values are normalized by the standard deviation of each tomographic channel. The generated maps visually reproduce the non-Gaussian and highly skewed structures present in the simulations, and preserve the strong correlations across tomographic channels. The third row compares the mean power spectrum and peak-count statistics between the simulations and generated samples. The curves show $(\mathrm{Flow}-\mathrm{True})/\sigma_{\mathrm{True}}$, 
where $\sigma_{\mathrm{True}}$ denotes the realization-to-realization scatter measured from the simulations. The gray band indicates agreement within $0.5\sigma$. The flow model accurately reproduces the average power spectra of all tomographic combinations and also captures the peak-count statistics. Since peak counts are a representative non-Gaussian observable beyond the two-point function, this agreement demonstrates that the model successfully learns non-Gaussian information in addition to the power spectrum. The fourth row shows the ratio of realization-to-realization scatter, $\sigma_{\mathrm{Flow}}/\sigma_{\mathrm{True}}$, for the same summary statistics. Although the mean statistics are reproduced accurately, the scatter is systematically suppressed for both power spectra and peak counts. This variance shrinkage motivates the latent corrections introduced in Section~\ref{subsec:correction}.
}
\end{figure*}

\subsection{Baseline flow model}

We employ a continuous normalizing flow (CNF) to model the probability distribution of weak-lensing convergence fields. Following the flow-matching framework \cite{Lipman2023}, the generated field is obtained by integrating a velocity field from a Gaussian initial state at $t=0$ to the target data distribution at $t=1$. The CNF evolves the convergence field according to 
\beqa 
\frac{\mathrm d\kappa}{\mathrm dt} = v(\bfx,z_s,t), 
\eeqa 
where $v$ denotes the learned velocity field.

To respect the statistical homogeneity and isotropy of cosmological fields, the velocity field is represented using isotropic spectral operators that act in Fourier space. The network combines scale-dependent filtering, channel mixing, and feature-wise linear modulation (FiLM) \cite{Perez2018}. Further implementation details are provided in Appendix~\ref{apdx:flow}.

The network is trained using the standard flow-matching objective \cite{Lipman2023}. 
Generated samples are obtained by integrating the learned flow dynamics from Gaussian initial conditions to the target weak-lensing distribution. 

Figure~\ref{fig:baseline_flow} summarizes the performance of the baseline model. The generated maps successfully reproduce the visual appearance of the simulations as well as the ensemble-average power spectra and peak-count statistics. However, the model systematically underestimates realization-to-realization scatter in both Gaussian and non-Gaussian summary statistics. This variance shrinkage motivates the physics-informed latent corrections introduced in the next section.

\subsection{Physics-informed latent corrections\label{subsec:correction}}

The baseline flow reproduces the average properties of the training simulations but exhibits variance shrinkage, leading to an underestimation of realization-to-realization fluctuations in weak-lensing observables. To recover the missing variability, we introduce latent corrections motivated by the coupling between long- and short-wavelength density fluctuations \cite{Takada:2013wfa,Bayer:2022nws}.

Weak-lensing observables measured in a finite field of view are sensitive to super-survey density modes outside the survey window \cite{Sato:2009ct,Barreira:2017fjz}. Through nonlinear gravitational evolution, these large-scale modes modulate the growth of small-scale structures and generate additional realization-to-realization scatter. Because our training data are constructed from random crops of larger simulations, different realizations effectively probe different large-scale environments. It is therefore natural to introduce latent variables that describe the leading departures from perfect homogeneity within each cropped field.

Rather than modeling arbitrary spatial fluctuations, 
we expand the latent field using only the monopole and the fundamental Fourier modes, 
\beqa 
\theta(\bfx) &=& \theta_0 +\theta_1^{\rm c}\cos(\ell_{0,1}x_1) +\theta_1^{\rm s}\sin(\ell_{0,1}x_1) \nonumber\\ && +\theta_2^{\rm c}\cos(\ell_{0,2}x_2) +\theta_2^{\rm s}\sin(\ell_{0,2}x_2), \eeqa 
where $\ell_{0,1}$ and $\ell_{0,2}$ denote the fundamental Fourier modes of the field. 
The latent coefficients are sampled from an isotropic Gaussian prior and are treated as realization-level nuisance variables. During the flow evolution, they remain fixed, 
\beqa 
\frac{\mathrm d\theta}{\mathrm dt}=0, 
\eeqa 
so that they modify the generative dynamics without introducing independent latent dynamics.

The latent variables affect the baseline velocity field through three physically motivated corrections. First, an additive correction 
\beqa 
v \rightarrow v+\sigma_{\rm add}\,t\,\tanh(\theta_{\rm add}) 
\eeqa 
introduces coherent shifts of the velocity field and increases the diversity of field means.

Second, the velocity field is modulated by a time-dependent multiplicative factor, 
\beqa 
v \rightarrow \alpha_t\,v, \eeqa with \beqa \alpha_t = \exp \!\left[ g(t) \left( \sigma_{\rm time}\theta_{\rm mul} + \sigma_{\rm add2}\sigma_{\rm add}\theta_{\rm add} \right) \right], \eeqa 
where 
\beqa g(t) = \left[ 1+\exp\!\left(-20(t-t_0)\right) \right]^{-1}. 
\eeqa 
In practice, this multiplicative term provides the dominant contribution to variance recovery and acts as a realization-dependent modulation of effective structure growth.

Finally, we introduce a nonlinear tail correction, 
\beqa 
v \rightarrow v + g(t)\, \sigma_{\rm tail}\, \theta_{\rm mul}^{+}\, D \, \frac{D^2}{1+D^2}, \eeqa 
where $\theta_{\rm mul}^{+}\equiv\max(\theta_{\rm mul},0)$ and $D$ is given by Eq.~(\ref{eq:flow_traject}). 
This term preferentially enhances the evolution of high-density regions and mimics the amplification of short-scale growth caused by nonlinear coupling between super-survey and local modes.

The correction parameters $\sigma_{\rm add}$, $\sigma_{\rm add2}$, $\sigma_{\rm time}$, 
$\sigma_{\rm tail}$, and $t_0$ are determined by matching the distributions of field means and field variances between generated and simulated maps on a channel-by-channel basis. 
The optimization process of the latent parameters is summarized in Appendix~\ref{apdx:latent_opt}.
As shown below, the dominant recovered latent direction behaves as an effective growth-amplitude mode.

\section{Recovering Missing Variability}

The baseline flow reproduces the average properties of the training simulations but exhibits variance shrinkage, leading to an underestimation of realization-to-realization scatter in weak-lensing observables. To recover the missing variability, we introduce the physics-informed latent corrections described in Section~\ref{subsec:correction}.

\begin{figure*}[!t] \includegraphics[clip, width=1.8\columnwidth]{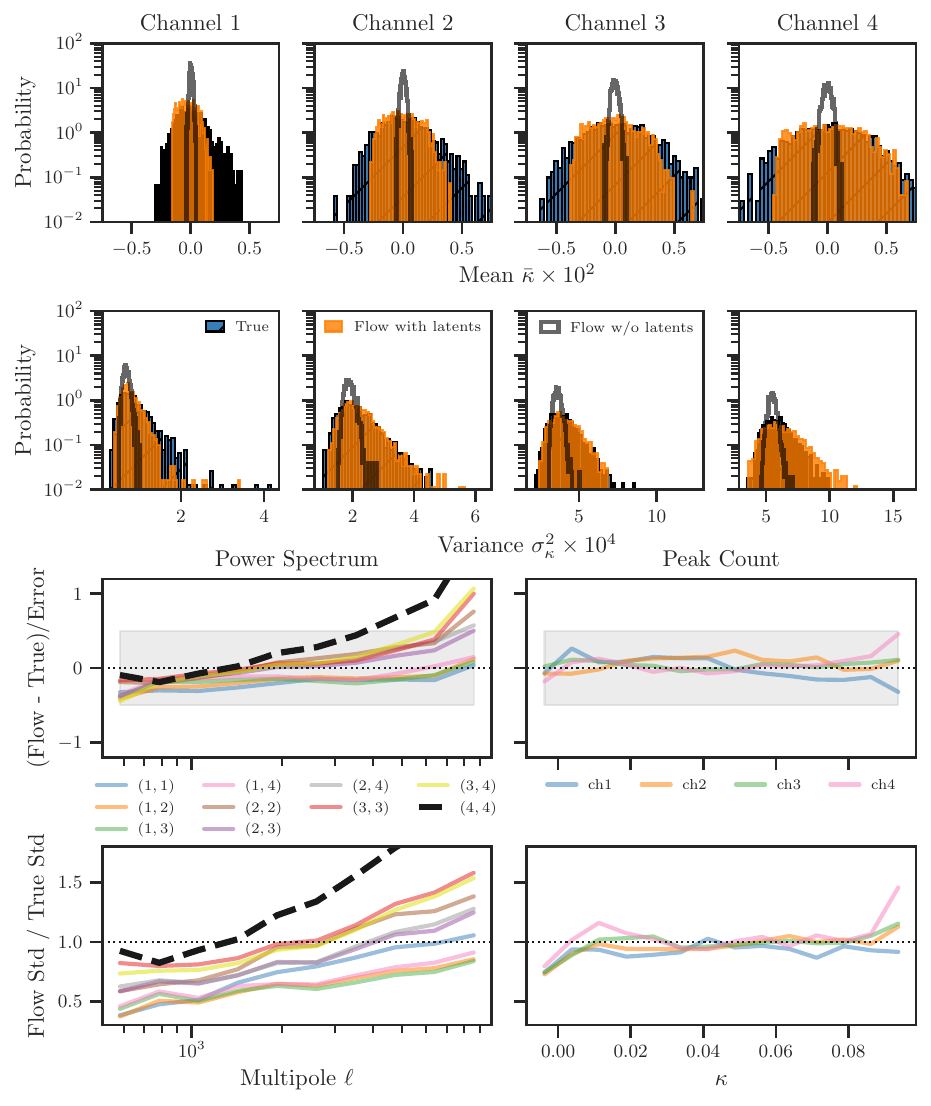} \caption{\label{fig:latent_flow} Performance of the latent-enhanced flow model. The first and second rows compare the distributions of the field mean and field variance for the four tomographic channels. Blue, orange, and gray histograms denote the simulations, the latent-enhanced flow model, and the baseline flow model, respectively. The third row compares the mean power spectra and peak-count statistics. As in Fig.~\ref{fig:baseline_flow}, the curves show $(\mathrm{Flow}-\mathrm{True})/\sigma_{\mathrm{True}}$, where $\sigma_{\mathrm{True}}$ denotes the realization-to-realization scatter measured from the simulations. The gray band indicates agreement within $0.5\sigma$. The fourth row shows the ratio of realization-to-realization scatter, $\sigma_{\mathrm{Flow}}/\sigma_{\mathrm{True}}$, for the same summary statistics. } \end{figure*}

Figure~\ref{fig:latent_flow} summarizes the performance of the latent-enhanced model. The latent corrections substantially broaden the distributions of both field means and field variances, bringing them into much better agreement with the simulations. In particular, the variance distributions, which are significantly underestimated by the baseline flow, are recovered in all tomographic channels.

Importantly, increasing the realization-to-realization variability does not significantly degrade the mean statistical predictions of the model. The mean power spectra remain within approximately $0.5\sigma$ of the simulations for most tomographic pairs, while the peak-count statistics are reproduced within roughly $0.5\sigma$ over the full range of peak heights. Since peak counts probe non-Gaussian information beyond the two-point function, this demonstrates that the latent corrections preserve the non-Gaussian structures learned by the baseline flow.

The impact of the latent corrections is most clearly seen in the scatter of summary statistics. Relative to the baseline model shown in Fig.~\ref{fig:baseline_flow}, the latent-enhanced flow exhibits substantially larger realization-to-realization variability. The improvement is particularly evident in the power-spectrum variance at $\ell\gtrsim3000$, where halo sample variance is expected to dominate \cite{Sato:2009ct}. An even stronger agreement is observed for peak counts, whose scatter is accurately reproduced over nearly the entire range of peak heights.

Taken together, these results indicate that the latent corrections recover a substantial fraction of the missing variability present in the training simulations while preserving the average weak-lensing observables. We therefore conclude that variance shrinkage in the baseline flow is largely associated with a small number of missing effective degrees of freedom that can be restored through physics-informed modifications of the generative dynamics.

\section{Response of Latent Variables\label{sec:response}}

\begin{figure*}[!t] \includegraphics[clip, width=1.8\columnwidth]{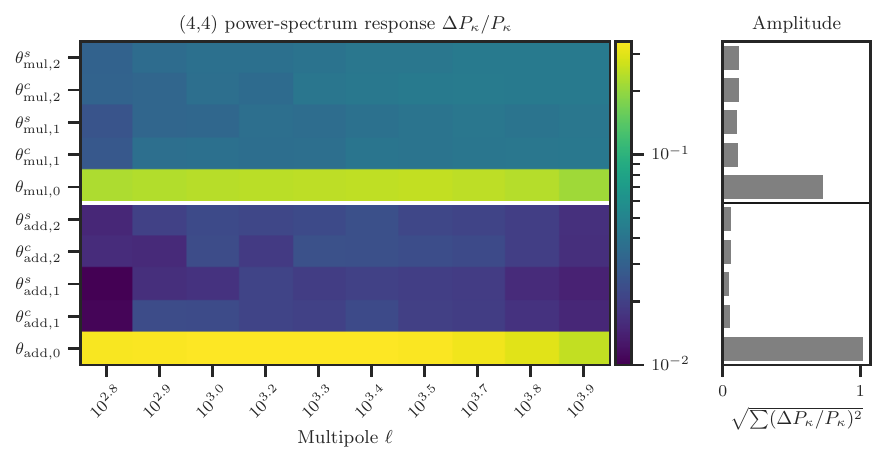} \caption{\label{fig:latent_pk_response} Fractional response of the $(4,4)$ convergence power spectrum to individual latent variables. For each latent parameter, the response is computed by setting the corresponding latent variable to unity while fixing all others to zero. The heat map shows the scale-dependent response, $\Delta P_\kappa/P_\kappa$, as a function of multipole and latent direction. The right panel shows the integrated response amplitude defined in Eq.~(\ref{eq:pk_response_norm}). Monopole latent modes dominate the response, with $\theta_{\rm add,0}$ producing the largest effect. } \end{figure*}

To identify the physical role of the latent variables, we measure the response of the convergence power spectrum to individual latent perturbations. For each latent parameter 
$\theta_i$, we set $\theta_i=1$ while fixing all remaining latent variables to zero and define the fractional response 
\beqa 
\frac{\Delta P_\kappa(\ell)} {P_\kappa(\ell)} = \frac{ P_\kappa(\ell;\theta_i=1) }{ P_\kappa(\ell;\theta_i=0) } -1. \label{eq:pk_response} 
\eeqa

Figure~\ref{fig:latent_pk_response} shows the resulting responses for the $(4,4)$ tomographic power spectrum. To quantify their overall importance, we define the integrated response amplitude 
\beqa 
A_i = \sqrt{ \sum_\ell \left[ \frac{\Delta P_\kappa(\ell)} {P_\kappa(\ell)} \right]^2 }. \label{eq:pk_response_norm} 
\eeqa

A clear hierarchy is visible among the latent directions. The dominant responses arise from the monopole modes $\theta_{\rm add,0}$ and $\theta_{\rm mul,0}$, with $\theta_{\rm add,0}$ providing the largest contribution. Higher-order spatial modes induce substantially weaker responses, indicating that the missing variability of the baseline flow is primarily associated with large-scale modulations of the generative dynamics. This observation motivates our focus on $\theta_{\rm add,0}$ throughout the remainder of the paper.

Importantly, the responses are not perfectly scale independent. Both the amplitude and shape of $\Delta P_\kappa/P_\kappa$ vary across multipoles, particularly for the dominant latent directions. Consequently, the latent corrections cannot be interpreted as simple global rescalings of the generated fields. Instead, they modify the generative dynamics itself and naturally produce scale-dependent changes in weak-lensing statistics.

Although Fig.~\ref{fig:latent_pk_response} shows only the $(4,4)$ tomographic power spectrum, the remaining tomographic combinations exhibit qualitatively similar behavior, with the monopole latent modes dominating the response and displaying comparable scale dependence.

\section{In-Distribution Field-Level Inference}

We assess the impact of the latent corrections through field-level inference of noiseless weak-lensing maps from noisy observations. Synthetic observations are generated by adding Gaussian noise to $\kappa$TNG realizations. The noise root-mean-square values are 0.034, 0.031, 0.034, and 0.049 for the four tomographic bins, respectively. 
The adopted noise level is approximately 25\% of that expected for the Subaru Hyper Suprime-Cam survey \cite{Li:2023tui}. We intentionally use a reduced noise level to isolate the impact of model uncertainty and to facilitate a controlled comparison between the fiducial and latent-enhanced priors. Since the rms of individual $\kappa$ realizations is typically $0.010$--$0.025$, the resulting observations are strongly noise dominated.

Posterior samples of noiseless $\kappa$ fields are obtained using the Metropolis-adjusted Langevin algorithm (MALA) implemented in BLACKJAX \cite{cabezas2024blackjax}. 
In our inference, posterior over latent variables $\bd{\theta}$ and $\kappa$ are sampled jointly, 
and $\bd{\theta}$ are marginalized rather than fixed.
Posterior samples are generated with eight independent MALA chains.
The full sampling configuration is summarized in Appendix~\ref{apdx:MALA}. 
The combined chains yield an effective sample size of approximately 600 
and satisfy $\hat R < 1.05$.
%We run eight independent chains with a step size of 0.003 and discard the first 30,000 samples as burn-in. Posterior samples are stored every 200 steps during an additional 30,000-step production run. After combining all chains, the effective sample size is approximately 600. We verified convergence using the Gelman--Rubin statistic computed from the posterior density, obtaining $\hat R<1.05$.

\begin{figure*}[!t]
\includegraphics[clip, width=1.8\columnwidth]{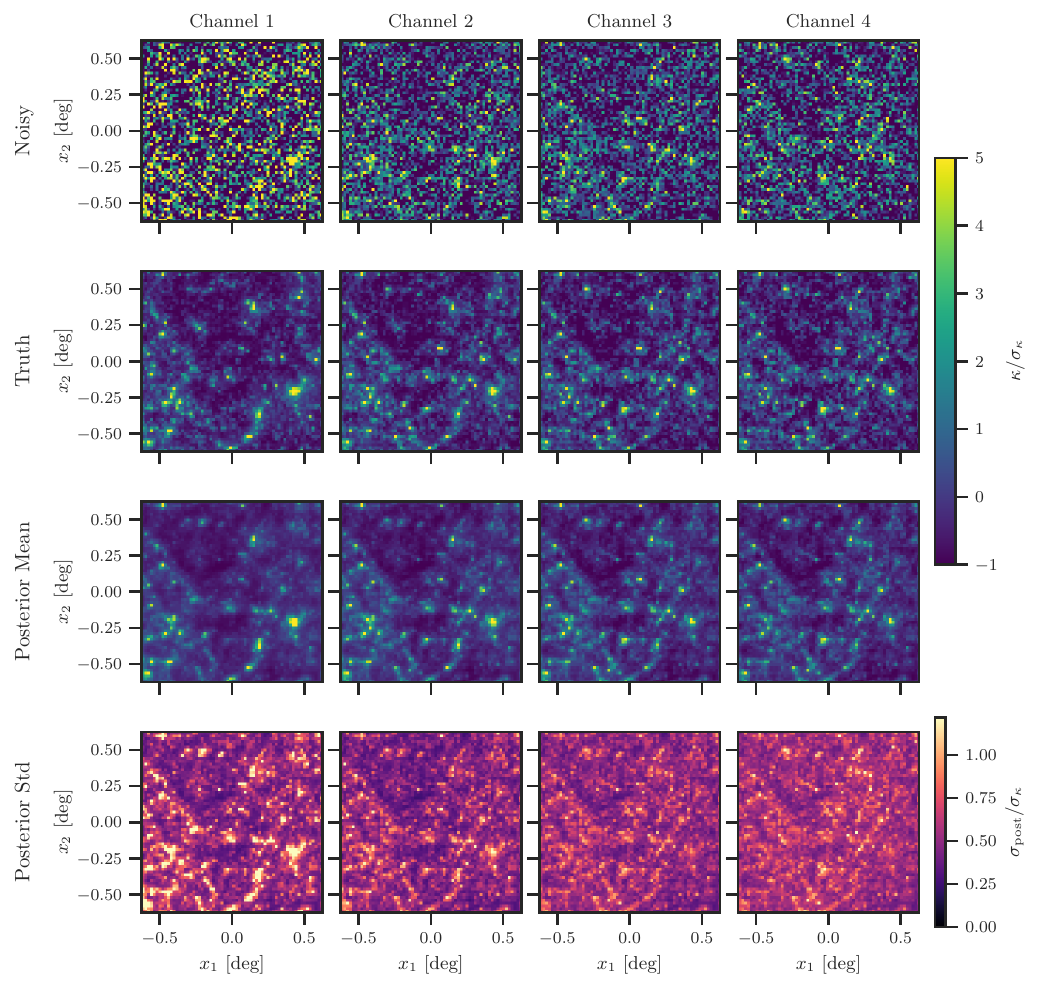}
\caption{\label{fig:FLI_example}
Example of field-level inference using the latent-enhanced flow prior. From top to bottom, the panels show the noisy observed convergence field, the underlying noiseless convergence field, the posterior mean reconstruction, and the posterior standard deviation. Columns correspond to the four tomographic source-redshift bins. Pixel values are normalized by the standard deviation of each tomographic channel. The first three rows share a common color scale, while the fourth row uses a separate color scale for the posterior standard deviation.
}
\end{figure*}

\subsubsection*{Visual comparisons}

We first validate the field-level inference pipeline using in-distribution data generated from the same simulation suite used for training. Although these test realizations are drawn from the same underlying simulation set, they are generated using independent random crops and random seeds and therefore do not appear in the training data. The resulting validation set thus provides a non-trivial test of the model's ability to generalize beyond the specific realizations used during training. Figure~\ref{fig:FLI_example} shows a representative reconstruction obtained with the latent-enhanced flow prior.

The first row displays the mock observed convergence maps after the addition of shape noise. At the adopted noise level, small-scale structures are largely obscured, making it difficult to visually identify the underlying matter distribution. The second row shows the corresponding noiseless convergence maps from the simulation. The third row presents the posterior mean reconstruction. Despite the substantial contamination by shape noise, the inferred convergences recover most of the prominent structures visible in the true maps. In particular, the strong correlations among tomographic channels are preserved by the reconstruction, indicating that the inference successfully exploits the shared large-scale structures across source-redshift bins.

The posterior mean reconstruction is visibly smoother than the true convergence field. 
This smoothing results from averaging over the posterior volume and is most apparent in high-density regions where multiple field realizations contribute to the posterior. Nevertheless, the dominant structures and their tomographic correlations remain well reproduced.

The fourth row shows the posterior standard deviation. Regions containing high-convergence peaks exhibit noticeably larger uncertainties than the surrounding low-density regions. This behavior arises because localized nonlinear structures cannot be reconstructed uniquely from the noisy observations: while their presence is strongly supported by the data, their exact amplitudes and positions remain uncertain within the posterior distribution.

The visual agreement between the posterior mean and the true convergence maps demonstrates that the latent-enhanced flow prior provides a realistic reconstruction of the underlying weak-lensing field for in-distribution observations. 

\subsubsection*{Summary statistics}

We next quantify the reconstruction accuracy using power-spectrum and peak-count statistics and compare the performance of baseline and latent-enhanced flow models.

\begin{figure*}[!t]
\includegraphics[clip, width=1.8\columnwidth]{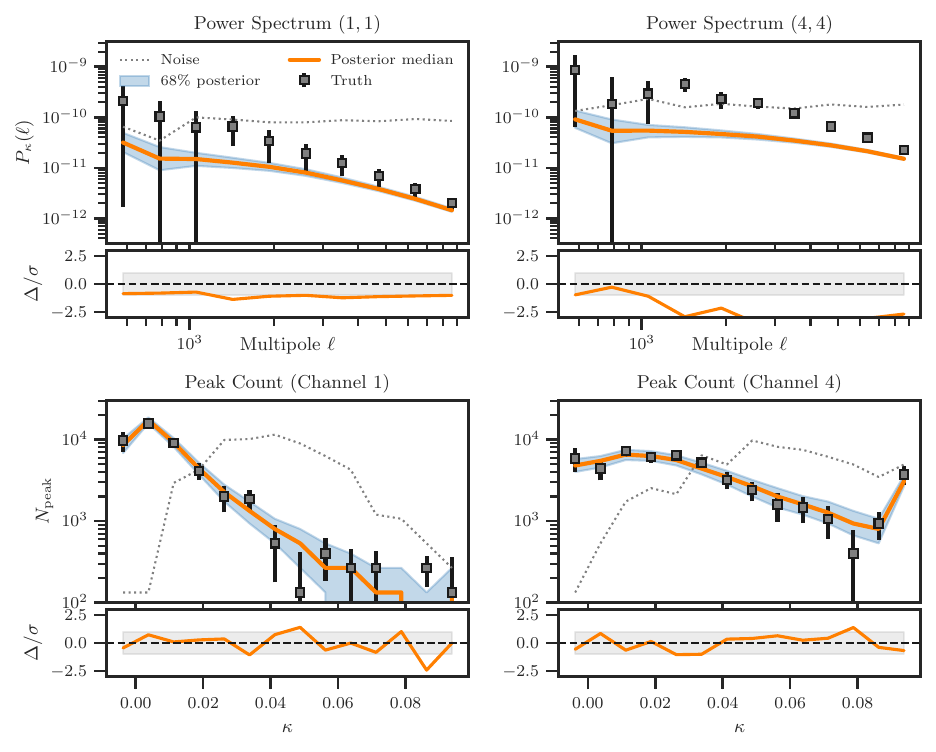}
\caption{\label{fig:ind_statistics}
Summary-statistic validation of the latent-enhanced flow model for an in-distribution mock observation. The upper panels compare the true noiseless statistics (black symbols), posterior median (orange), 68\% posterior credible interval (blue band), and the corresponding noise-only statistics (gray dotted curves). Black error bars indicate the realization-to-realization scatter of the noiseless convergence field. The lower panels show the normalized residuals, $(\mathrm{Truth}-\mathrm{Posterior~Median})/\sigma$, where $\sigma$ denotes the noiseless scatter. The gray bands correspond to the 
$\pm1\sigma$ region.
}
\end{figure*}

Figure~\ref{fig:ind_statistics} shows the inferred summary statistics for a representative in-distribution realization. The upper panels compare the true noiseless statistics with the posterior predictive distributions obtained from the latent-enhanced flow prior. The posterior median is shown by the orange curves, the 68\% posterior credible intervals are indicated by the blue shaded regions, and the gray dotted lines correspond to the statistics measured from noise-only realizations. The black error bars denote the realization-to-realization scatter of the noiseless convergence field in the simulation ensemble. The lower panels show the normalized residuals, 
\beqa 
\Delta/\sigma = \frac{ {\rm Truth} - {\rm Posterior~Median} }{ \sigma_{\rm noiseless} }, 
\eeqa 
where $\sigma_{\rm noiseless}$ is the standard deviation of the noiseless convergence-field statistics.

The first row shows the power-spectrum reconstruction for the $(1,1)$ and $(4,4)$ tomographic power spectra. The posterior median power spectrum, computed from the ensemble of posterior samples, is systematically lower than the true power spectrum, particularly at high multipoles. This trend persists when replacing the posterior median with the posterior mean, indicating that it is not driven by an asymmetry of the posterior distribution. 
This discrepancy does not originate from the flow prior itself, which accurately reproduces the power spectra of the training simulations (Fig.~\ref{fig:baseline_flow}). Rather, it appears only after conditioning on the noisy observations through field-level inference.

For the $(1,1)$ auto-spectrum, the recovered power remains within approximately $1\sigma$ of the truth even in the noise-dominated regime. The $(4,4)$ spectrum exhibits deviations exceeding $2\sigma$ at $\ell\gtrsim1500$. These results suggest that the likelihood provides only limited constraints on the smallest angular scales, causing the posterior to favor somewhat smoother realizations than the true convergence field. It should be emphasized, however, that the noiseless scatter adopted here provides a particularly stringent performance metric. In practical weak-lensing analyses, power-spectrum uncertainties are usually estimated from noisy realizations and are therefore substantially larger than the error bars shown in this figure.

The second row presents the corresponding peak-count statistics. Unlike the power spectrum, the posterior median closely follows the true noiseless peak counts over the entire range of peak heights. The reconstruction does not collapse toward the noise-only prediction despite the significant shape-noise contamination present in the observed maps. Instead, the inferred peak-count distributions remain consistent with the underlying noiseless realization at roughly the $1\sigma$ level. Since peak counts probe highly non-Gaussian structures associated with nonlinear gravitational collapse, this agreement indicates that the field-level inference accurately recovers the dominant non-Gaussian features of the convergence field.
This agreement is particularly notable because peak counts are not explicitly used during inference and therefore provide an independent validation of the reconstructed non-Gaussian structures.

%Overall, these results demonstrate that the latent-enhanced flow prior successfully reconstructs both Gaussian and non-Gaussian summary statistics in the in-distribution setting. 
%We note that the posterior power spectra show a mild suppression at intermediate multipoles despite accurately reproducing the field variance and peak-count statistics. This effect appears only in the field-level inference and not in unconditional samples from the flow prior. A detailed investigation of its origin is beyond the scope of the present work.
Overall, these results establish that the latent-enhanced flow provides a reliable in-distribution field-level prior, forming the basis for the out-of-distribution analyses presented below.

\begin{figure*}[!t]
\includegraphics[clip, width=2.\columnwidth]{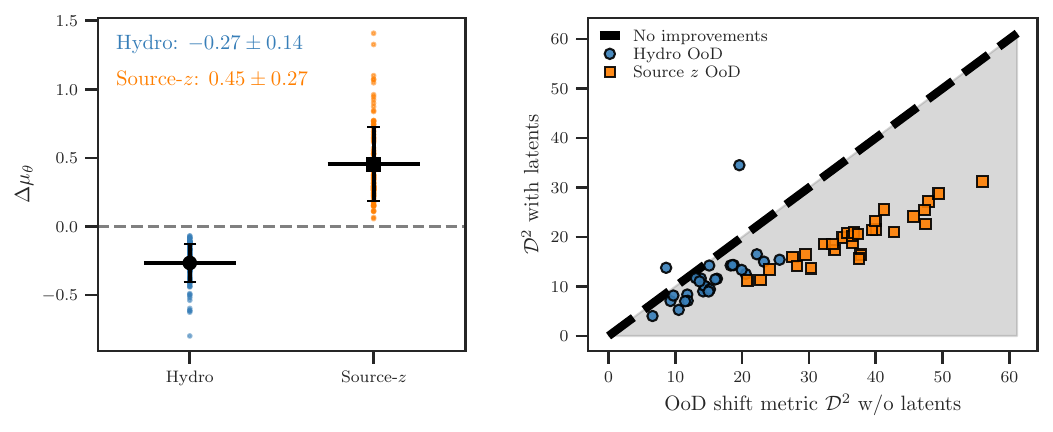}
\caption{\label{fig:OoD_summary} 
Left: Shift of the inferred latent posterior mean between out-of-distribution (OoD) and in-distribution (inD) analyses, 
$\Delta \mu_\theta \equiv \mu(\theta_{\rm add,0}|{\rm OoD})-\mu(\theta_{\rm add,0}|{\rm inD})$. Hydro OoD realizations systematically shift the latent variable toward negative values, whereas source-redshift OoD realizations shift it toward positive values. This demonstrates that the latent degree of freedom is actively utilized during field-level inference.
Right: OoD shift metric ${\cal D}^2$ obtained with and without latent degrees of freedom. Points below the diagonal correspond to improved agreement with the target OoD distribution. 
The latent-enhanced model consistently reduces the discrepancy for both hydro and source-redshift shifts, indicating improved robustness to model mismatch.
}
\end{figure*}

\section{Out-of-Distribution Robustness}

\subsection{OoD setup}

To evaluate the robustness of field-level inference against distribution shifts, we consider two types of out-of-distribution (OoD) data. The first, referred to as the \emph{Hydro OoD} case, is generated from the full-physics simulations while keeping the source-redshift configuration identical to that used for training. This setup isolates the impact of baryonic physics that is absent from the gravity-only simulations used to train the generative model.

The second, referred to as the \emph{Source-$z$ OoD} case, uses the same gravity-only simulations as the training data but adopts slightly higher source redshifts in all tomographic channels. The source-redshift configuration is 
\beqa (z_s^{(1)},z_s^{(2)},z_s^{(3)},z_s^{(4)}) = (0.552,\,0.803,\,1.097,\,1.375), 
\eeqa 
corresponding to shifts of approximately $0.05$--$0.07$ relative to the fiducial setup. These shifts are comparable to typical photometric-redshift uncertainties in current weak-lensing analyses.

No retraining is performed in either OoD test. Instead, we apply the field-level inference pipeline using the model trained exclusively on the in-distribution simulations. The observational noise model, likelihood function, MALA configuration, and all other inference settings are kept identical to those adopted in the in-distribution analyses. This setup provides a controlled test of whether the latent corrections can improve robustness against realistic astrophysical and observational distribution shifts.

\subsection{Posterior latent response}

We first investigate how the latent variables are utilized during field-level inference. For each realization, we compare the posterior mean of the dominant latent mode between the in-distribution and out-of-distribution analyses, 
\beqa \Delta\mu_{\theta} \equiv \mu(\theta_{\rm add,0}|{\rm OoD}) - \mu(\theta_{\rm add,0}|{\rm inD}). 
\eeqa

As shown in the left panel of Fig.~\ref{fig:OoD_summary}, hydrodynamical realizations systematically shift $\theta_{\rm add,0}$ toward negative values, whereas source-redshift shifts induce positive posterior shifts. These trends are consistent with the power-spectrum responses shown in Fig.~\ref{fig:latent_pk_response}: baryonic effects suppress the lensing power relative to the gravity-only training set, while increasing the source redshift enhances the projected lensing signal.

Most importantly, the sign of $\Delta\mu_{\theta}$ is coherent across all realizations. The dominant latent variable therefore responds systematically to the direction of the underlying OoD shift rather than fluctuating randomly. This provides direct evidence that the latent degree of freedom is actively exploited during inference to absorb part of the discrepancy between the training and target distributions.

\subsection{OoD metric}

To quantify the impact of the latent corrections on OoD inference, we define an OoD shift metric 
\begin{equation} 
\begin{split} 
{\cal D}^2 =& \sum_i \frac{ (R_i-\hat R_i)^2 } { {\rm Var}(R_i) }, \\ 
R_i =& \log \left( P_i^{\rm true,OoD} / P_i^{\rm true,inD} \right), \\ 
\hat R_i =& \log \left( P_i^{\rm post,OoD} / P_i^{\rm post,inD} \right). 
\end{split}
\end{equation}

Here the index $i$ runs over all tomographic pairs and ten logarithmic multipole bins in the range $500\le\ell\le10^4$. The metric compares the true OoD response with that inferred from the posterior distribution. The reconstructed power spectrum is defined as the median of the power spectra evaluated from posterior samples.

Because each OoD realization is generated from the same initial phases as its in-distribution counterpart, the comparison can be performed on a realization-by-realization basis. The right panel of Fig.~\ref{fig:OoD_summary} shows ${\cal D}^{2}$ obtained with and without latent corrections. Points below the diagonal correspond to improved agreement with the true OoD response.

For both the Hydro OoD and Source-$z$ OoD tests, the latent-enhanced model systematically moves the inferred power-spectrum response toward the true OoD direction, producing smaller values of ${\cal D}^{2}$ for most realizations. The latent variables therefore do more than simply increase the variance of generated samples; they generate physically meaningful responses that improve robustness against distribution shifts.

Examples of the power-spectrum responses that enter the definition of ${\cal D}^2$ are presented in Appendix~\ref{apdx:OoD_extra}, together with the corresponding peak-count responses.

Taken together, these results suggest that variance shrinkage in simulation-trained generative models can be interpreted as evidence of missing effective degrees of freedom. By introducing physics-informed latent corrections to the generative dynamics, a substantial fraction of the missing variability can be recovered and subsequently utilized during field-level inference. The resulting model displays improved robustness to realistic distribution shifts without requiring additional training simulations.

\section{Interpretation as an Effective Amplitude Degree of Freedom}

To gain physical insight into the dominant latent direction, we compare the latent-enhanced flow model with a simple amplitude-nuisance extension of the baseline flow. The goal is to determine whether the dominant latent mode, $\theta_{\rm add,0}$, behaves as an effective amplitude degree of freedom during field-level inference.

Specifically, we consider a modified flow model of the form \beqa 
\kappa_{\rm model} = \alpha\, \kappa_{\rm flow}, 
\eeqa 
where $\kappa_{\rm flow}$ denotes the convergence field generated by the baseline flow model and $\alpha$ is a multiplicative amplitude parameter. 
We allow an independent amplitude parameter 
for each tomographic channel. 
Since $\alpha$ rescales the convergence field uniformly, 
this model generates a scale-independent modification of the power spectrum, 
\beqa P_\kappa(\ell) \rightarrow \alpha^2 P_\kappa(\ell), 
\eeqa 
and therefore provides a useful baseline model 
for capturing the lowest-order component of OoD shifts.
%is therefore expected to absorb the lowest-order component of OoD shifts. 

\begin{figure}[!t]
\includegraphics[clip, width=1\columnwidth]{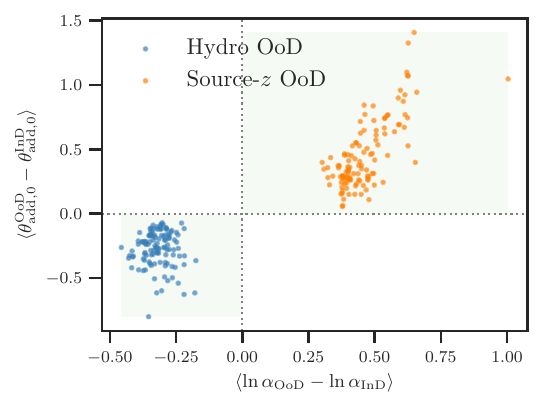}
\caption{ \label{fig:alpha_theta} Comparison between the OoD-induced shifts of the effective amplitude parameter and the dominant latent variable. The horizontal axis shows the posterior-mean shift in $\log\alpha$ inferred using the amplitude-nuisance model, while the vertical axis shows the posterior-mean shift in $\theta_{\rm add,0}$ inferred using the latent-enhanced model. Points correspond to all realizations and tomographic channels for the Hydro OoD (blue) and Source-$z$ OoD (orange) tests. }
\end{figure}

Field-level inference requires a prior on the amplitude parameters. We adopt a lognormal prior and assume that 
\beqa 
\log \alpha \sim \mathcal N(0,\sigma_\alpha^2), 
\eeqa 
with the amplitude parameters 
in different tomographic channels treated as independent. 
The prior width $\sigma_\alpha$ is calibrated directly 
from the fiducial gravity-only simulations used to train the generative model. 
For each realization, 
we estimate an effective amplitude parameter through 
\beqa 
\log \hat{\alpha} = \frac{ \sum_i \frac{1}{2} \log \left[ \frac{ P(\ell_i) }{ \langle P(\ell_i)\rangle } \right] }{ \sum_i 1 }, 
\eeqa 
where $P(\ell_i)$ is the power spectrum of an individual realization and $\langle P(\ell_i)\rangle$ 
denotes the ensemble-averaged power spectrum of the training set. 
The standard deviation of $\log \hat{\alpha}$ over realizations is then adopted as the prior width. 
Applying this procedure to the fiducial gravity-only simulations yields $\sigma_\alpha = (0.1998,\, 0.1529,\, 0.1281,\, 0.1139)$ for the first through fourth tomographic channels, respectively. These values define the Gaussian priors used for the amplitude-nuisance inference presented below.

We apply the amplitude-nuisance model described above to the same field-level inference framework used throughout this work. In addition to the convergence field, the four amplitude parameters $\alpha$ are sampled simultaneously with the flow degrees of freedom using the MALA sampler. The Gaussian likelihood, noise model, and all sampling settings are identical to those adopted in the baseline and latent-enhanced analyses. For each realization, we compute the posterior mean amplitudes in both the in-distribution and OoD cases and define the OoD-induced amplitude shift through 
\beqa 
\Delta \log\alpha = \langle \log\alpha_{\rm OoD} \rangle - \langle \log\alpha_{\rm InD} \rangle. 
\eeqa 
We then compare these amplitude shifts with the corresponding shifts inferred for the dominant latent variable $\theta_{\rm add,0}$ in the latent-enhanced model. 
%If the dominant latent mode acts as an effective amplitude degree of freedom, the OoD-induced shifts inferred from the two approaches should exhibit a consistent response.
If the dominant latent mode indeed acts as an effective amplitude degree of freedom, the inferred shifts $\Delta\log\alpha$ and $\Delta\theta_{\rm add,0}$ should respond coherently to the OoD perturbation.

Figure~\ref{fig:alpha_theta} shows 
the posterior-mean shifts inferred from the two approaches. 
The horizontal axis corresponds to the shift in the effective amplitude parameter, while the vertical axis shows the corresponding shift in the dominant latent variable, 
$\langle \theta_{\rm add,0}^{\rm OoD} - \theta_{\rm add,0}^{\rm InD} \rangle$.
Each point corresponds to one realization and one tomographic channel. We combine all four tomographic channels and all 30 realizations for both OoD tests.

%A clear separation between the two OoD scenarios is observed. 
The two OoD scenarios occupy distinct quadrants of Fig.~\ref{fig:alpha_theta}.
For the Hydro OoD case, both the effective amplitude parameter and the dominant latent variable shift toward negative values. This behavior is expected because baryonic feedback suppresses the small-scale lensing power spectrum relative to the gravity-only simulations used for training. In contrast, the Source-$z$ OoD case produces positive shifts in both quantities, consistent with the increase in lensing amplitude generated by the higher source-redshift distribution.

%Most importantly, all realizations lie either in the first quadrant or in the third quadrant. In other words, the inferred shifts in $\theta_{\rm add,0}$ always have the same sign as the shifts inferred for the effective amplitude parameter. This behavior would be highly unlikely if the latent-enhanced model were responding randomly to OoD data. Instead, the result indicates that the latent variables consistently identify the direction of the discrepancy between the OoD observations and the in-distribution model and respond by shifting the posterior toward the expected OoD direction.
Most importantly, all realizations lie either in the first or third quadrant, implying that the OoD-induced shifts of $\theta_{\rm add,0}$ and $\log\alpha$ always have the same sign. The latent-enhanced model therefore responds coherently to the underlying OoD perturbation rather than generating random fluctuations in latent space.

The existence of a coherent sign relationship does not imply that $\theta_{\rm add,0}$ is identical to the amplitude parameter $\alpha$. The amplitude-nuisance model generates a purely scale-independent rescaling of the power spectrum, whereas the latent response shown in Section~\ref{sec:response} is generally scale dependent. 
Nevertheless, the strong alignment between the OoD responses demonstrates that the dominant latent variable absorbs, at least partially, the effective amplitude uncertainty of the generative model. This provides a natural physical interpretation for the improved OoD robustness of the latent-enhanced field-level inference framework.
%This explains why the latent-enhanced field-level inference is able to produce OoD responses that are systematically closer to the simulation truth than those obtained with the latent-free baseline model.

\section{Discussion and Conclusions}

In this paper, we have demonstrated that physics-informed latent corrections can recover a substantial fraction of the realization-level variability lost in simulation-trained generative models. The recovered latent degree of freedom is not only statistically meaningful but also actively utilized during field-level inference, leading to improved robustness against distribution shifts arising from both baryonic physics and source-redshift mismatches.

More broadly, our results suggest a general framework for improving simulation-trained generative models. Rather than treating variance shrinkage solely as a limitation of finite training data, it can be interpreted as a diagnostic of missing effective degrees of freedom in the learned generative dynamics. The missing degrees of freedom may then be identified through statistical discrepancies between simulations and model predictions and incorporated through physically motivated modifications of the generative process.

In the present work, we focused on variance shrinkage and found that the dominant recovered latent direction behaves as an effective growth-amplitude nuisance parameter. Importantly, this interpretation was not imposed a priori but emerged from a generic latent correction designed to recover missing variability. Our framework therefore provides a way to identify missing effective degrees of freedom directly from discrepancies between simulations and model predictions. By comparing the latent-enhanced model with an explicit amplitude-nuisance parameterization, we showed that OoD-induced shifts in the dominant latent variable closely track those inferred for the effective amplitude parameter. The latent corrections therefore recover a physically meaningful response direction directly from the training data, rather than requiring such a nuisance parameter to be specified beforehand.

Unlike a phenomenological amplitude parameter, however, the latent correction acts on the generative dynamics itself and naturally produces scale-dependent power-spectrum responses, as shown in Fig.~\ref{fig:latent_pk_response}. This suggests that the latent-enhanced flow provides a more flexible description of model uncertainty than a simple global rescaling of generated fields.

However, the covariance structure of weak-lensing observables is only partially recovered by the current model, indicating that additional latent directions may exist beyond the dominant amplitude mode. Determining which latent corrections are required to reproduce higher-order statistics and nontrivial covariance structures remains an open question. Since the latent corrections act directly on the generative dynamics, theoretical frameworks such as effective field theory may provide physically motivated guidance for constructing and interpreting such degrees of freedom.

Although our implementation is based on a continuous normalizing flow, the underlying idea naturally extends to diffusion and score-based models, whose generation processes are likewise governed by differential equations. In contrast, incorporating interpretable latent degrees of freedom into feed-forward neural-network emulators remains considerably less straightforward, despite their computational efficiency. Developing such capabilities may therefore be an important direction for future scientific machine-learning applications.

Our results highlight a complementary relationship between physics and deep learning. Deep generative models efficiently learn the average behavior encoded in finite simulation datasets, whereas physics-informed latent dynamics can recover part of the variability that is difficult to learn from data alone. We therefore anticipate that diagnosing variance shrinkage and restoring missing degrees of freedom through modifications of generative dynamics will provide a broadly applicable strategy for simulation-based field-level inference and scientific generative modeling.

\section*{Data Availability} The weak-lensing convergence maps used in this work are derived from the publicly available $\kappa$TNG simulations. All data products required to reproduce the analyses can be generated from these simulations following the procedures described in the paper. The trained flow-model weights, latent-enhanced model checkpoints, and additional inference products are available from the corresponding author upon reasonable request.

\begin{acknowledgments}
MS is supported by JSPS KAKENHI Grant Numbers JP24H00221 and JP24H00215. 
MS also acknowledges research supports by JST BOOST, Japan Grant Number JP-MJBY24D8.
\end{acknowledgments}

\appendix

\section{Weak-lensing observables\label{apdx:lensing}} 

Weak gravitational lensing is the distortion of images of distant galaxies by the gravitational potential of foreground large-scale structures. 
In the weak-lensing regime, the relevant projected density field is the convergence field $\kappa$, 
which can be expressed as a weighted line-of-sight projection of the matter overdensity field \cite{Bartelmann:1999yn}, 
\beqa 
\kappa(\bfx) = \int d\chi \, W(\chi)\, \delta_{\rm m}(\chi,\bfx), 
\eeqa 
where $\bfx$ denotes the angular position, $\chi$ is the comoving radial distance, $\delta_{\rm m}$ is the matter overdensity field, 
and $W(\chi)$ is the lensing efficiency kernel determined by the source-redshift distribution. 
Because $\kappa$ traces the projected matter distribution, its statistical properties provide a powerful probe of cosmic structure formation. 

The primary summary statistic considered in this work is the flat-sky convergence power spectrum. 
Defining the Fourier transform of the convergence field as 
\beqa 
\tilde{\kappa}(\boldsymbol{\ell}) = \int d^2\bfx\, \kappa(\boldsymbol{\theta}) e^{-i\boldsymbol{\ell}\cdot \bfx}, 
\eeqa 
the power spectrum is given by 
\beqa 
\left\langle \tilde{\kappa}(\boldsymbol{\ell}) \tilde{\kappa}^*(\boldsymbol{\ell}') \right\rangle = (2\pi)^2 \delta_\mathrm{D} (\boldsymbol{\ell}-\boldsymbol{\ell}') P_\kappa(\ell), 
\eeqa 
where $\delta_\mathrm{D}$ is the Dirac delta function and $\ell=|\boldsymbol{\ell}|$. 
The power spectrum captures the Gaussian information of the convergence field. Throughout this work, it is measured over the range $500\le\ell\le10^4$ using 10 logarithmically spaced bins.

To probe non-Gaussian information, we additionally consider peak counts. A peak is defined as a pixel whose convergence value exceeds those of its eight neighboring pixels. Peak counts are measured in 14 linearly spaced bins over the range $-0.0075 \le \kappa \le 0.0975$, roughly correponding to 
$-0.5\sigma$ to $6.5\sigma$-level peaks in our simulations. Since peak counts are sensitive to nonlinear structure formation and massive halos, they provide complementary information beyond that contained in the power spectrum.

\section{Training data preprocessing\label{apdx:preprocess}}

To remove large-scale modes that cannot be supported within the smaller training patches, we apply a Fourier-space filter 
\beqa 
W(\ell)=\exp\left[-\left(\frac{\ell_{\rm cut}}{\ell}\right)^2\right]. 
\eeqa
The cutoff scale $\ell_{\rm cut}$ is chosen such that the variance of filtered and randomly cropped $5\times5\,\mathrm{deg}^2$ Gaussian fields matches that of Gaussian fields generated directly in a $1.25\times1.25\,\mathrm{deg}^2$ box. 
The Gaussian fields are generated 
using the mean power spectrum of the gravity-only $\kappa$TNG simulations at the corresponding source redshift. 
This procedure approximately removes super-window fluctuations at the Gaussian level. 
The resulting values of $\ell_{\rm cut}$ are 393, 336, 367, and 425 
for increasing source redshift.

We then apply the log-normal transformation 
\beqa 
y(\bfx, z_s) = \ln\left[1+\kappa(\bfx, z_s)/q\right], 
\eeqa 
where 
%$z_s$ is the source redshift, and 
$q$ sets an effective minimum value of $\kappa$. 
We define $q=|\mathrm{MIN}[\kappa(\bfx,z_s)]|+0.001$ using all 10000 realizations. 
The transformed fields are subsequently standardized to improve training stability. 
The final training target consists of four-channel convergence maps with $64\times64$ pixels, where the channels remain strongly correlated through the common underlying large-scale structure.

\section{Flow architecture\label{apdx:flow}}

We employ a continuous normalizing flow (CNF) to generate four-channel convergence maps. 
The CNF learns a velocity field that transports samples from an isotropic Gaussian prior to the target non-Gaussian distribution. 
Following the flow-matching framework \cite{Lipman2023}, 
we define a straight-line probability path between Gaussian noise and data, 
\beqa 
D(\bfx, z_s, t) = (1-t)u_\mathrm{G}(\bfx, z_s) + t\tilde{y}(\bfx, z_s), \label{eq:flow_traject} \eeqa 
where $u_\mathrm{G}$ is a Gaussian white-noise realization and $\tilde{y}$ denotes 
the log-normal transformed and standardized convergence field. 
The corresponding target velocity is 
\beqa 
v(\bfx,z_s,t) = \tilde{y}(\bfx,z_s)-u_\mathrm{G}(\bfx,z_s). 
\eeqa

We parameterize the velocity field with a neural network $v_\phi(\bfx,z_s,t)$, where $\phi$ denotes the trainable parameters. To respect the statistical isotropy and homogeneity of cosmological fields, the network is built from spectral operators that depend only on the norm of the Fourier mode.

The input field is lifted to 32 feature channels, processed through six residual blocks, and projected back to four output channels. Each block consists of 
\begin{itemize} 
\item a spectral operator, 
\item a pointwise linear transformation, 
\item FiLM conditioning \cite{Perez2018}, 
\item a GeLU activation \cite{Hendrycks2016}. 
\end{itemize}
The spectral operator acts in Fourier space and performs both scale-dependent filtering and channel mixing. The channel-mixing operation enables interactions among different spectral features and is intended to capture the effective mode couplings generated by nonlinear gravitational evolution.

Specifically, the Fourier kernel is expanded in a basis comprising 24 Gaussian filters centered at uniformly spaced Fourier scales together with a Lorentzian low-pass component, 
$\left[1+\left(2\ell/\ell_\mathrm{f}\right)^2\right]^{-1}$, 
where $\ell_\mathrm{f}=288$ corresponds to the fundamental mode of the $1.25\times1.25\,\mathrm{deg}^2$ field of view. The spectral operator is represented as a weighted sum of these basis functions, with trainable channel-mixing matrices as coefficients. The resulting Fourier-space features are transformed back to real space and combined with a local pointwise residual branch. To improve training stability, the spatial mean of the output velocity field is removed at the end of each residual block.

The flow-time dependence is modeled through FiLM conditioning. 
The flow time $t$ is encoded using a 16-dimensional sinusoidal embedding and mapped to a 32-dimensional latent vector $\bd{e}$ through a two-layer MLP. Each residual block is then conditioned according to 
$\bd{\gamma}(\bd{e})\odot\bd{h}+\bd{\beta}(\bd{e})$, 
where $\bd{h}$ denotes the block activations and $\bd{\gamma}$ and $\bd{\beta}$ are learned linear projections of $\bd e$. The resulting network contains 532900 trainable parameters.

The network parameters are trained using the flow-matching objective \cite{Lipman2023}, 
which minimizes the discrepancy between the neural-network velocity field $v_\phi$ and the target velocity defined along the interpolation path of Eq.~(\ref{eq:flow_traject}). In the end, we optimize the mean-squared error 
\beqa 
\mathcal{L}_{\rm FM} = \left\langle \left| v_\phi(\bfx,z_s,t) - v(\bfx,z_s,t) \right|^2 \right\rangle, 
\eeqa 
where the average is taken over training fields and randomly sampled flow times $t\sim U(0,1)$.
We train the network using the Adam optimizer with a learning rate of $10^{-3}$ and a batch size of 100. Training is performed for 1500 optimization steps. We confirmed that the flow-matching loss reaches a clear plateau by the end of training, indicating convergence of the optimization.

Generated maps are obtained by integrating the learned flow dynamics from Gaussian initial conditions at $t=0$ to the target distribution at $t=1$. Unless otherwise stated, the ODE is solved using a first-order Euler method with 50 steps.

\section{Latent optimization\label{apdx:latent_opt}}

After training the baseline flow, all neural-network parameters are held fixed. We then optimize only the parameters governing the latent corrections described in Section~\ref{subsec:correction}.

The optimization objective is designed to recover the realization-to-realization variability lost during training. In particular, we match the distributions of field means and field variances between generated maps and training simulations. The field kurtosis is included as a weak regularization term to suppress unphysical solutions.

For a given minibatch of simulated maps, latent-corrected samples are generated by integrating the flow dynamics. We then compute the mean, variance, and kurtosis of each realization and construct a differentiable soft-histogram loss based on the first two moments of their distributions, 
\beqa 
\mathcal{L}_\theta = \mathcal{L}_{\rm mean} + \mathcal{L}_{\rm var} + 0.25\,\mathcal{L}_{\rm kurt}, 
\eeqa 
where the mean and variance terms dominate the optimization. The loss compares the average and scatter of the generated and simulated realizations in each minibatch.

The latent parameters are optimized using the Adam optimizer with a learning rate of $10^{-2}$ and a batch size of 40. 
We find that the optimization converges rapidly, with most of the improvement occurring within the first $\sim10$ optimization steps.

Model selection is performed independently from the training loss. 
After every optimization step, we evaluate the agreement between generated and simulated maps using the one-dimensional Wasserstein distance of the field-mean, field-variance, and field-kurtosis distributions. The model yielding the smallest validation metric, 
\beqa 
\mathcal{M} = W_{\rm mean} + W_{\rm var} + 0.25\,W_{\rm kurt}, 
\eeqa 
is retained as the final latent-corrected model. We verified that the resulting latent corrections substantially improve the distributions of field means and variances while preserving the higher-order properties of the generated weak-lensing maps.

\section{MALA details\label{apdx:MALA}}

To assess the impact of latent corrections on field-level inference, we generate mock observations from the $\kappa$TNG simulations. Following the same preprocessing as used for training, we first apply the large-scale filtering described in Section~\ref{apdx:preprocess} and then randomly crop $1.25\times1.25\,{\rm deg}^2$ patches. Shape noise is subsequently added to each tomographic channel assuming an intrinsic ellipticity dispersion of $\sigma_e=0.26$ and galaxy number densities of $(64,\,80,\,64,\,32)\,{\rm arcmin}^{-2}$. The resulting noise level is approximately $25\%$ of that expected for the HSC survey and is chosen to test denoising performance down to small angular scales. The goal of the inference is to reconstruct the underlying noiseless convergence field from these noisy observations.

We perform Bayesian field-level inference using 
the Metropolis-adjusted Langevin algorithm (MALA) \cite{Besag1994,Roberts1996} 
as implemented in \texttt{BlackJAX} \cite{BlackJAX}. 
For a state vector $\mathbf{q}$ containing the convergence field and latent variables, 
the proposal distribution is 
\beqa 
\mathbf{q}' = \mathbf{q} + \frac{\epsilon^2}{2} \nabla_{\mathbf{q}} \log p(\mathbf{q}|d) + \epsilon\,\boldsymbol{\eta}, 
\eeqa 
where $\boldsymbol{\eta}\sim\mathcal N(0,\mathbf{I})$ and 
$\epsilon$ is the MALA step size. Throughout this work we adopt $\epsilon=0.003$.

The posterior distribution is constructed from a Gaussian likelihood and the flow prior, 
\beqa 
p(\mathbf{q}|d) \propto p(d|\mathbf{q}) \,p_{\rm flow}(\mathbf{q}), 
\eeqa 
where the flow prior includes the latent-corrected velocity field introduced in the main text. The corresponding CNF log-density is evaluated by integrating the divergence term along the flow trajectory, which is computed numerically using the Hutchinson trace estimator.

All analyses are performed using eight independent Markov chains. 
The initial latent variables are set to zero. To initialize the flow degrees of freedom, 
we randomly select convergence maps from the $\kappa$TNG training set 
and integrate the flow dynamics backward from $t=1$ to $t=0$ 
to obtain the corresponding latent representation. 
Each chain is evolved for 30,000 burn-in samples followed by 30,000 production samples. 
During the production phase, posterior convergence maps are stored every 200 samples.

Convergence is monitored using the Gelman--Rubin statistic computed from the posterior density. For all analyses presented in this work, we obtain $\hat R<1.05$. The combined chains yield an effective sample size of approximately 600 posterior samples.

\section{Additional OoD examples\label{apdx:OoD_extra}}

\begin{figure*}[!t]
\includegraphics[clip, width=1.8\columnwidth]{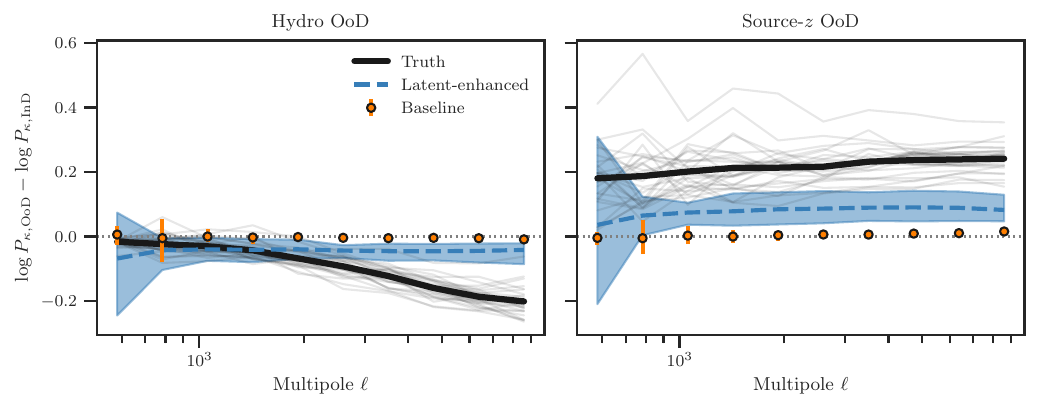}
\caption{\label{fig:ood_shift_pk} Power-spectrum responses to out-of-distribution (OoD) shifts for the Hydro OoD (left) and Source-$z$ OoD (right) tests. The plotted quantity is the logarithmic difference between the OoD and in-distribution auto power spectra in the first tomographic channel. Thin black curves show realization-level responses measured from the simulations, while the thick black curve shows the mean response over 30 realizations. The blue dashed curve and shaded region denote the mean, minimum, and maximum responses inferred with the latent-enhanced model. Orange points and error bars show the corresponding predictions from the latent-free baseline model.
}
\end{figure*}

\begin{figure*}[!t]
\includegraphics[clip, width=1.8\columnwidth]{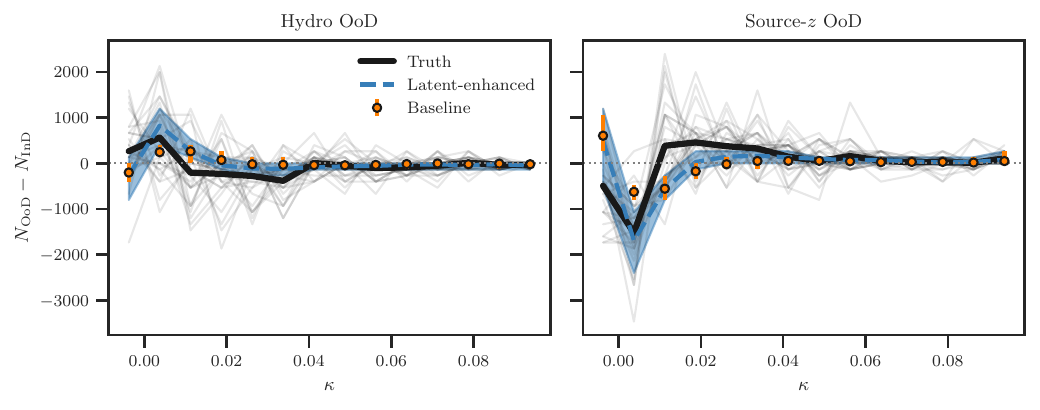}
\caption{\label{fig:ood_shift_peak} Same as Fig.~\ref{fig:ood_shift_pk}, but showing the OoD response of the peak-count statistic in the first tomographic channel.
}
\end{figure*}

The main text quantifies OoD robustness using the ensemble-level metric ${\cal D}^2$. Here we directly examine the inferred responses of individual summary statistics by comparing paired OoD and in-distribution realizations generated from identical initial phases.

Figure~\ref{fig:ood_shift_pk} shows the logarithmic power-spectrum responses for the Hydro OoD and Source-$z$ OoD tests. The simulation responses exhibit substantial realization-to-realization scatter, indicating that OoD effects are intrinsically realization dependent. Nevertheless, the mean responses display clear trends. In the Hydro OoD case, baryonic effects suppress small-scale power and generate a negative response that increases toward higher multipoles. In contrast, increasing the source redshift enhances the lensing signal and produces a positive response over most multipoles.

The latent-free baseline model captures only a small fraction of these shifts. Its inferred responses remain close to zero and exhibit little realization-to-realization scatter. By comparison, the latent-enhanced model produces substantially broader responses and consistently moves in the direction of the true OoD signal. Although the amplitudes of the inferred responses remain smaller than those measured directly from the simulations, the latent variables recover a significant fraction of the variability absent in the baseline model.

Figure~\ref{fig:ood_shift_peak} presents the corresponding analysis using peak counts. The qualitative conclusions remain unchanged. The true OoD responses exhibit considerable realization dependence, while the baseline model predicts weaker and less diverse shifts. The latent-enhanced model generates both larger responses and increased realization-to-realization scatter, bringing the inferred peak-count statistics closer to the simulation results.

Taken together, these results support the interpretation proposed in the main text. The latent corrections do not merely modify the mean prediction of the generative model; they also recover part of the realization-level variability required to describe out-of-distribution data.

% The \nocite command causes all entries in a bibliography to be printed out
% whether or not they are actually referenced in the text. This is appropriate
% for the sample file to show the different styles of references, but authors
% most likely will not want to use it.
%\nocite{*}

\bibliography{refs}% Produces the bibliography via BibTeX.

%apsrev4-2.bst 2019-01-14 (MD) hand-edited version of apsrev4-1.bst
%Control: key (0)
%Control: author (8) initials jnrlst
%Control: editor formatted (1) identically to author
%Control: production of article title (0) allowed
%Control: page (0) single
%Control: year (1) truncated
%Control: production of eprint (0) enabled
\begin{thebibliography}{23}%
\makeatletter
\providecommand \@ifxundefined [1]{%
 \@ifx{#1\undefined}
}%
\providecommand \@ifnum [1]{%
 \ifnum #1\expandafter \@firstoftwo
 \else \expandafter \@secondoftwo
 \fi
}%
\providecommand \@ifx [1]{%
 \ifx #1\expandafter \@firstoftwo
 \else \expandafter \@secondoftwo
 \fi
}%
\providecommand \natexlab [1]{#1}%
\providecommand \enquote  [1]{``#1''}%
\providecommand \bibnamefont  [1]{#1}%
\providecommand \bibfnamefont [1]{#1}%
\providecommand \citenamefont [1]{#1}%
\providecommand \href@noop [0]{\@secondoftwo}%
\providecommand \href [0]{\begingroup \@sanitize@url \@href}%
\providecommand \@href[1]{\@@startlink{#1}\@@href}%
\providecommand \@@href[1]{\endgroup#1\@@endlink}%
\providecommand \@sanitize@url [0]{\catcode `\\12\catcode `\$12\catcode `\&12\catcode `\#12\catcode `\^12\catcode `\_12\catcode `\%12\relax}%
\providecommand \@@startlink[1]{}%
\providecommand \@@endlink[0]{}%
\providecommand \url  [0]{\begingroup\@sanitize@url \@url }%
\providecommand \@url [1]{\endgroup\@href {#1}{\urlprefix }}%
\providecommand \urlprefix  [0]{URL }%
\providecommand \Eprint [0]{\href }%
\providecommand \doibase [0]{https://doi.org/}%
\providecommand \selectlanguage [0]{\@gobble}%
\providecommand \bibinfo  [0]{\@secondoftwo}%
\providecommand \bibfield  [0]{\@secondoftwo}%
\providecommand \translation [1]{[#1]}%
\providecommand \BibitemOpen [0]{}%
\providecommand \bibitemStop [0]{}%
\providecommand \bibitemNoStop [0]{.\EOS\space}%
\providecommand \EOS [0]{\spacefactor3000\relax}%
\providecommand \BibitemShut  [1]{\csname bibitem#1\endcsname}%
\let\auto@bib@innerbib\@empty
%</preamble>
\bibitem [{\citenamefont {Jasche}\ and\ \citenamefont {Wandelt}(2013)}]{Jasche:2012kq}%
  \BibitemOpen
  \bibfield  {author} {\bibinfo {author} {\bibfnamefont {J.}~\bibnamefont {Jasche}}\ and\ \bibinfo {author} {\bibfnamefont {B.~D.}\ \bibnamefont {Wandelt}},\ }\bibfield  {title} {\bibinfo {title} {{Bayesian physical reconstruction of initial conditions from large scale structure surveys}},\ }\href {https://doi.org/10.1093/mnras/stt449} {\bibfield  {journal} {\bibinfo  {journal} {Mon. Not. Roy. Astron. Soc.}\ }\textbf {\bibinfo {volume} {432}},\ \bibinfo {pages} {894} (\bibinfo {year} {2013})},\ \Eprint {https://arxiv.org/abs/1203.3639} {arXiv:1203.3639 [astro-ph.CO]} \BibitemShut {NoStop}%
\bibitem [{\citenamefont {Schmidt}\ \emph {et~al.}(2020)\citenamefont {Schmidt}, \citenamefont {Cabass}, \citenamefont {Jasche},\ and\ \citenamefont {Lavaux}}]{Schmidt:2020viy}%
  \BibitemOpen
  \bibfield  {author} {\bibinfo {author} {\bibfnamefont {F.}~\bibnamefont {Schmidt}}, \bibinfo {author} {\bibfnamefont {G.}~\bibnamefont {Cabass}}, \bibinfo {author} {\bibfnamefont {J.}~\bibnamefont {Jasche}},\ and\ \bibinfo {author} {\bibfnamefont {G.}~\bibnamefont {Lavaux}},\ }\bibfield  {title} {\bibinfo {title} {{Unbiased Cosmology Inference from Biased Tracers using the EFT Likelihood}},\ }\href {https://doi.org/10.1088/1475-7516/2020/11/008} {\bibfield  {journal} {\bibinfo  {journal} {JCAP}\ }\textbf {\bibinfo {volume} {11}},\ \bibinfo {pages} {008}},\ \Eprint {https://arxiv.org/abs/2004.06707} {arXiv:2004.06707 [astro-ph.CO]} \BibitemShut {NoStop}%
\bibitem [{\citenamefont {Leclercq}(2025)}]{Leclercq2025}%
  \BibitemOpen
  \bibfield  {author} {\bibinfo {author} {\bibfnamefont {F.}~\bibnamefont {Leclercq}},\ }\bibfield  {title} {\bibinfo {title} {Field-level inference in cosmology},\ }\href@noop {} {\  (\bibinfo {year} {2025})},\ \Eprint {https://arxiv.org/abs/2509.13435} {arXiv:2509.13435 [astro-ph.CO]} \BibitemShut {NoStop}%
\bibitem [{\citenamefont {Shirasaki}\ and\ \citenamefont {Ikeda}(2024)}]{Shirasaki:2023nnk}%
  \BibitemOpen
  \bibfield  {author} {\bibinfo {author} {\bibfnamefont {M.}~\bibnamefont {Shirasaki}}\ and\ \bibinfo {author} {\bibfnamefont {S.}~\bibnamefont {Ikeda}},\ }\bibfield  {title} {\bibinfo {title} {{Neural style transfer of weak lensing mass maps}},\ }\href {https://doi.org/10.33232/001c.118104} {\bibfield  {journal} {\bibinfo  {journal} {The Open Journal of Astrophysics}\ }\textbf {\bibinfo {volume} {7}},\ \bibinfo {pages} {42} (\bibinfo {year} {2024})},\ \Eprint {https://arxiv.org/abs/2310.17141} {arXiv:2310.17141 [astro-ph.CO]} \BibitemShut {NoStop}%
\bibitem [{\citenamefont {Armijo}\ \emph {et~al.}(2026)\citenamefont {Armijo}, \citenamefont {Thiele},\ and\ \citenamefont {Liu}}]{Armijo:2026ucb}%
  \BibitemOpen
  \bibfield  {author} {\bibinfo {author} {\bibfnamefont {J.}~\bibnamefont {Armijo}}, \bibinfo {author} {\bibfnamefont {L.}~\bibnamefont {Thiele}},\ and\ \bibinfo {author} {\bibfnamefont {J.}~\bibnamefont {Liu}},\ }\bibfield  {title} {\bibinfo {title} {{Replicating weak-lensing summary-statistic covariances with normalizing flows}},\ }\href {https://doi.org/10.1103/lbb5-2kbt} {\bibfield  {journal} {\bibinfo  {journal} {Phys. Rev. D}\ }\textbf {\bibinfo {volume} {114}},\ \bibinfo {pages} {023567} (\bibinfo {year} {2026})},\ \Eprint {https://arxiv.org/abs/2601.20669} {arXiv:2601.20669 [astro-ph.CO]} \BibitemShut {NoStop}%
\bibitem [{\citenamefont {Doeser}\ and\ \citenamefont {Jasche}(2026)}]{Doeser:2026gnj}%
  \BibitemOpen
  \bibfield  {author} {\bibinfo {author} {\bibfnamefont {L.}~\bibnamefont {Doeser}}\ and\ \bibinfo {author} {\bibfnamefont {J.}~\bibnamefont {Jasche}},\ }\bibfield  {title} {\bibinfo {title} {{Learning the Universe: Posterior Reliability of Neural Generative Models in High-Dimensional Field-Level Inference of Cosmic Initial Conditions}},\ }\href@noop {} {\  (\bibinfo {year} {2026})},\ \Eprint {https://arxiv.org/abs/2606.10023} {arXiv:2606.10023 [astro-ph.CO]} \BibitemShut {NoStop}%
\bibitem [{\citenamefont {Thiele}(2026)}]{Thiele2026}%
  \BibitemOpen
  \bibfield  {author} {\bibinfo {author} {\bibfnamefont {L.}~\bibnamefont {Thiele}},\ }\bibfield  {title} {\bibinfo {title} {Machine learning techniques for astrophysics and cosmology: Simulation-based inference},\ }\href@noop {} {\  (\bibinfo {year} {2026})},\ \Eprint {https://arxiv.org/abs/2605.10719} {arXiv:2605.10719 [astro-ph.CO]} \BibitemShut {NoStop}%
\bibitem [{\citenamefont {Rau}\ \emph {et~al.}(2023)\citenamefont {Rau}, \citenamefont {Dalal}, \citenamefont {Zhang}, \citenamefont {Li}, \citenamefont {Nishizawa}, \citenamefont {More}, \citenamefont {Mandelbaum}, \citenamefont {Miyatake}, \citenamefont {Strauss},\ and\ \citenamefont {Takada}}]{Rau:2022wrq}%
  \BibitemOpen
  \bibfield  {author} {\bibinfo {author} {\bibfnamefont {M.~M.}\ \bibnamefont {Rau}}, \bibinfo {author} {\bibfnamefont {R.}~\bibnamefont {Dalal}}, \bibinfo {author} {\bibfnamefont {T.}~\bibnamefont {Zhang}}, \bibinfo {author} {\bibfnamefont {X.}~\bibnamefont {Li}}, \bibinfo {author} {\bibfnamefont {A.~J.}\ \bibnamefont {Nishizawa}}, \bibinfo {author} {\bibfnamefont {S.}~\bibnamefont {More}}, \bibinfo {author} {\bibfnamefont {R.}~\bibnamefont {Mandelbaum}}, \bibinfo {author} {\bibfnamefont {H.}~\bibnamefont {Miyatake}}, \bibinfo {author} {\bibfnamefont {M.~A.}\ \bibnamefont {Strauss}},\ and\ \bibinfo {author} {\bibfnamefont {M.}~\bibnamefont {Takada}},\ }\bibfield  {title} {\bibinfo {title} {{Weak lensing tomographic redshift distribution inference for the Hyper Suprime-Cam Subaru Strategic Program three-year shape catalogue}},\ }\href {https://doi.org/10.1093/mnras/stad1962} {\bibfield  {journal} {\bibinfo  {journal} {Mon. Not. Roy. Astron. Soc.}\ }\textbf {\bibinfo {volume} {524}},\ \bibinfo {pages} {5109}
  (\bibinfo {year} {2023})},\ \Eprint {https://arxiv.org/abs/2211.16516} {arXiv:2211.16516 [astro-ph.CO]} \BibitemShut {NoStop}%
\bibitem [{\citenamefont {Osato}\ \emph {et~al.}(2021)\citenamefont {Osato}, \citenamefont {Liu},\ and\ \citenamefont {Haiman}}]{Osato:2020sxo}%
  \BibitemOpen
  \bibfield  {author} {\bibinfo {author} {\bibfnamefont {K.}~\bibnamefont {Osato}}, \bibinfo {author} {\bibfnamefont {J.}~\bibnamefont {Liu}},\ and\ \bibinfo {author} {\bibfnamefont {Z.}~\bibnamefont {Haiman}},\ }\bibfield  {title} {\bibinfo {title} {{$\kappa$TNG: effect of baryonic processes on weak lensing with IllustrisTNG simulations}},\ }\href {https://doi.org/10.1093/mnras/stab395} {\bibfield  {journal} {\bibinfo  {journal} {Mon. Not. Roy. Astron. Soc.}\ }\textbf {\bibinfo {volume} {502}},\ \bibinfo {pages} {5593} (\bibinfo {year} {2021})},\ \Eprint {https://arxiv.org/abs/2010.09731} {arXiv:2010.09731 [astro-ph.CO]} \BibitemShut {NoStop}%
\bibitem [{\citenamefont {Pillepich}\ \emph {et~al.}(2018)\citenamefont {Pillepich} \emph {et~al.}}]{Pillepich:2017fcc}%
  \BibitemOpen
  \bibfield  {author} {\bibinfo {author} {\bibfnamefont {A.}~\bibnamefont {Pillepich}} \emph {et~al.},\ }\bibfield  {title} {\bibinfo {title} {{First results from the IllustrisTNG simulations: the stellar mass content of groups and clusters of galaxies}},\ }\href {https://doi.org/10.1093/mnras/stx3112} {\bibfield  {journal} {\bibinfo  {journal} {Mon. Not. Roy. Astron. Soc.}\ }\textbf {\bibinfo {volume} {475}},\ \bibinfo {pages} {648} (\bibinfo {year} {2018})},\ \Eprint {https://arxiv.org/abs/1707.03406} {arXiv:1707.03406 [astro-ph.GA]} \BibitemShut {NoStop}%
\bibitem [{\citenamefont {Lipman}\ \emph {et~al.}(2023)\citenamefont {Lipman}, \citenamefont {Chen}, \citenamefont {Ben-Hamu}, \citenamefont {Nickel},\ and\ \citenamefont {Le}}]{Lipman2023}%
  \BibitemOpen
  \bibfield  {author} {\bibinfo {author} {\bibfnamefont {Y.}~\bibnamefont {Lipman}}, \bibinfo {author} {\bibfnamefont {R.~T.~Q.}\ \bibnamefont {Chen}}, \bibinfo {author} {\bibfnamefont {H.}~\bibnamefont {Ben-Hamu}}, \bibinfo {author} {\bibfnamefont {M.}~\bibnamefont {Nickel}},\ and\ \bibinfo {author} {\bibfnamefont {M.}~\bibnamefont {Le}},\ }\bibfield  {title} {\bibinfo {title} {Flow matching for generative modeling},\ }\href@noop {} {\  (\bibinfo {year} {2023})},\ \Eprint {https://arxiv.org/abs/2210.02747} {arXiv:2210.02747 [cs.LG]} \BibitemShut {NoStop}%
\bibitem [{\citenamefont {Perez}\ \emph {et~al.}(2018)\citenamefont {Perez}, \citenamefont {Strub}, \citenamefont {de~Vries}, \citenamefont {Dumoulin},\ and\ \citenamefont {Courville}}]{Perez2018}%
  \BibitemOpen
  \bibfield  {author} {\bibinfo {author} {\bibfnamefont {E.}~\bibnamefont {Perez}}, \bibinfo {author} {\bibfnamefont {F.}~\bibnamefont {Strub}}, \bibinfo {author} {\bibfnamefont {H.}~\bibnamefont {de~Vries}}, \bibinfo {author} {\bibfnamefont {V.}~\bibnamefont {Dumoulin}},\ and\ \bibinfo {author} {\bibfnamefont {A.}~\bibnamefont {Courville}},\ }\bibfield  {title} {\bibinfo {title} {{FiLM}: Visual reasoning with a general conditioning layer},\ }in\ \href@noop {} {\emph {\bibinfo {booktitle} {Proceedings of the AAAI Conference on Artificial Intelligence}}},\ Vol.~\bibinfo {volume} {32}\ (\bibinfo {year} {2018})\ \Eprint {https://arxiv.org/abs/1709.07871} {arXiv:1709.07871 [cs.LG]} \BibitemShut {NoStop}%
\bibitem [{\citenamefont {Takada}\ and\ \citenamefont {Hu}(2013)}]{Takada:2013wfa}%
  \BibitemOpen
  \bibfield  {author} {\bibinfo {author} {\bibfnamefont {M.}~\bibnamefont {Takada}}\ and\ \bibinfo {author} {\bibfnamefont {W.}~\bibnamefont {Hu}},\ }\bibfield  {title} {\bibinfo {title} {{Power Spectrum Super-Sample Covariance}},\ }\href {https://doi.org/10.1103/PhysRevD.87.123504} {\bibfield  {journal} {\bibinfo  {journal} {Phys. Rev. D}\ }\textbf {\bibinfo {volume} {87}},\ \bibinfo {pages} {123504} (\bibinfo {year} {2013})},\ \Eprint {https://arxiv.org/abs/1302.6994} {arXiv:1302.6994 [astro-ph.CO]} \BibitemShut {NoStop}%
\bibitem [{\citenamefont {Bayer}\ \emph {et~al.}(2023)\citenamefont {Bayer}, \citenamefont {Liu}, \citenamefont {Terasawa}, \citenamefont {Barreira}, \citenamefont {Zhong},\ and\ \citenamefont {Feng}}]{Bayer:2022nws}%
  \BibitemOpen
  \bibfield  {author} {\bibinfo {author} {\bibfnamefont {A.~E.}\ \bibnamefont {Bayer}}, \bibinfo {author} {\bibfnamefont {J.}~\bibnamefont {Liu}}, \bibinfo {author} {\bibfnamefont {R.}~\bibnamefont {Terasawa}}, \bibinfo {author} {\bibfnamefont {A.}~\bibnamefont {Barreira}}, \bibinfo {author} {\bibfnamefont {Y.}~\bibnamefont {Zhong}},\ and\ \bibinfo {author} {\bibfnamefont {Y.}~\bibnamefont {Feng}},\ }\bibfield  {title} {\bibinfo {title} {{Super-sample covariance of the power spectrum, bispectrum, halos, voids, and their cross covariances}},\ }\href {https://doi.org/10.1103/PhysRevD.108.043521} {\bibfield  {journal} {\bibinfo  {journal} {Phys. Rev. D}\ }\textbf {\bibinfo {volume} {108}},\ \bibinfo {pages} {043521} (\bibinfo {year} {2023})},\ \Eprint {https://arxiv.org/abs/2210.15647} {arXiv:2210.15647 [astro-ph.CO]} \BibitemShut {NoStop}%
\bibitem [{\citenamefont {Sato}\ \emph {et~al.}(2009)\citenamefont {Sato}, \citenamefont {Hamana}, \citenamefont {Takahashi}, \citenamefont {Takada}, \citenamefont {Yoshida}, \citenamefont {Matsubara},\ and\ \citenamefont {Sugiyama}}]{Sato:2009ct}%
  \BibitemOpen
  \bibfield  {author} {\bibinfo {author} {\bibfnamefont {M.}~\bibnamefont {Sato}}, \bibinfo {author} {\bibfnamefont {T.}~\bibnamefont {Hamana}}, \bibinfo {author} {\bibfnamefont {R.}~\bibnamefont {Takahashi}}, \bibinfo {author} {\bibfnamefont {M.}~\bibnamefont {Takada}}, \bibinfo {author} {\bibfnamefont {N.}~\bibnamefont {Yoshida}}, \bibinfo {author} {\bibfnamefont {T.}~\bibnamefont {Matsubara}},\ and\ \bibinfo {author} {\bibfnamefont {N.}~\bibnamefont {Sugiyama}},\ }\bibfield  {title} {\bibinfo {title} {{Simulations of Wide-Field Weak Lensing Surveys I: Basic Statistics and Non-Gaussian Effects}},\ }\href {https://doi.org/10.1088/0004-637X/701/2/945} {\bibfield  {journal} {\bibinfo  {journal} {Astrophys. J.}\ }\textbf {\bibinfo {volume} {701}},\ \bibinfo {pages} {945} (\bibinfo {year} {2009})},\ \Eprint {https://arxiv.org/abs/0906.2237} {arXiv:0906.2237 [astro-ph.CO]} \BibitemShut {NoStop}%
\bibitem [{\citenamefont {Barreira}\ \emph {et~al.}(2018)\citenamefont {Barreira}, \citenamefont {Krause},\ and\ \citenamefont {Schmidt}}]{Barreira:2017fjz}%
  \BibitemOpen
  \bibfield  {author} {\bibinfo {author} {\bibfnamefont {A.}~\bibnamefont {Barreira}}, \bibinfo {author} {\bibfnamefont {E.}~\bibnamefont {Krause}},\ and\ \bibinfo {author} {\bibfnamefont {F.}~\bibnamefont {Schmidt}},\ }\bibfield  {title} {\bibinfo {title} {{Complete super-sample lensing covariance in the response approach}},\ }\href {https://doi.org/10.1088/1475-7516/2018/06/015} {\bibfield  {journal} {\bibinfo  {journal} {JCAP}\ }\textbf {\bibinfo {volume} {06}},\ \bibinfo {pages} {015}},\ \Eprint {https://arxiv.org/abs/1711.07467} {arXiv:1711.07467 [astro-ph.CO]} \BibitemShut {NoStop}%
\bibitem [{\citenamefont {Li}\ \emph {et~al.}(2023)\citenamefont {Li} \emph {et~al.}}]{Li:2023tui}%
  \BibitemOpen
  \bibfield  {author} {\bibinfo {author} {\bibfnamefont {X.}~\bibnamefont {Li}} \emph {et~al.},\ }\bibfield  {title} {\bibinfo {title} {{Hyper Suprime-Cam Year 3 results: Cosmology from cosmic shear two-point correlation functions}},\ }\href {https://doi.org/10.1103/PhysRevD.108.123518} {\bibfield  {journal} {\bibinfo  {journal} {Phys. Rev. D}\ }\textbf {\bibinfo {volume} {108}},\ \bibinfo {pages} {123518} (\bibinfo {year} {2023})},\ \Eprint {https://arxiv.org/abs/2304.00702} {arXiv:2304.00702 [astro-ph.CO]} \BibitemShut {NoStop}%
\bibitem [{\citenamefont {Cabezas}\ \emph {et~al.}(2024)\citenamefont {Cabezas}, \citenamefont {Corenflos}, \citenamefont {Lao},\ and\ \citenamefont {Louf}}]{cabezas2024blackjax}%
  \BibitemOpen
  \bibfield  {author} {\bibinfo {author} {\bibfnamefont {A.}~\bibnamefont {Cabezas}}, \bibinfo {author} {\bibfnamefont {A.}~\bibnamefont {Corenflos}}, \bibinfo {author} {\bibfnamefont {J.}~\bibnamefont {Lao}},\ and\ \bibinfo {author} {\bibfnamefont {R.}~\bibnamefont {Louf}},\ }\href@noop {} {\bibinfo {title} {Blackjax: Composable {B}ayesian inference in {JAX}}} (\bibinfo {year} {2024}),\ \Eprint {https://arxiv.org/abs/2402.10797} {arXiv:2402.10797 [cs.MS]} \BibitemShut {NoStop}%
\bibitem [{\citenamefont {Bartelmann}\ and\ \citenamefont {Schneider}(2001)}]{Bartelmann:1999yn}%
  \BibitemOpen
  \bibfield  {author} {\bibinfo {author} {\bibfnamefont {M.}~\bibnamefont {Bartelmann}}\ and\ \bibinfo {author} {\bibfnamefont {P.}~\bibnamefont {Schneider}},\ }\bibfield  {title} {\bibinfo {title} {{Weak gravitational lensing}},\ }\href {https://doi.org/10.1016/S0370-1573(00)00082-X} {\bibfield  {journal} {\bibinfo  {journal} {Phys. Rept.}\ }\textbf {\bibinfo {volume} {340}},\ \bibinfo {pages} {291} (\bibinfo {year} {2001})},\ \Eprint {https://arxiv.org/abs/astro-ph/9912508} {arXiv:astro-ph/9912508} \BibitemShut {NoStop}%
\bibitem [{\citenamefont {Hendrycks}\ and\ \citenamefont {Gimpel}(2016)}]{Hendrycks2016}%
  \BibitemOpen
  \bibfield  {author} {\bibinfo {author} {\bibfnamefont {D.}~\bibnamefont {Hendrycks}}\ and\ \bibinfo {author} {\bibfnamefont {K.}~\bibnamefont {Gimpel}},\ }\bibfield  {title} {\bibinfo {title} {Gaussian error linear units (gelus)},\ }\href@noop {} {\  (\bibinfo {year} {2016})},\ \Eprint {https://arxiv.org/abs/1606.08415} {arXiv:1606.08415 [cs.LG]} \BibitemShut {NoStop}%
\bibitem [{\citenamefont {Besag}(1994)}]{Besag1994}%
  \BibitemOpen
  \bibfield  {author} {\bibinfo {author} {\bibfnamefont {J.}~\bibnamefont {Besag}},\ }\bibfield  {title} {\bibinfo {title} {Comments on ``representations of knowledge in complex systems'' by u. grenander and m. i. miller},\ }\href@noop {} {\bibfield  {journal} {\bibinfo  {journal} {Journal of the Royal Statistical Society: Series B}\ }\textbf {\bibinfo {volume} {56}},\ \bibinfo {pages} {591} (\bibinfo {year} {1994})}\BibitemShut {NoStop}%
\bibitem [{\citenamefont {Roberts}\ \emph {et~al.}(1996)\citenamefont {Roberts}, \citenamefont {Gelman},\ and\ \citenamefont {Gilks}}]{Roberts1996}%
  \BibitemOpen
  \bibfield  {author} {\bibinfo {author} {\bibfnamefont {G.~O.}\ \bibnamefont {Roberts}}, \bibinfo {author} {\bibfnamefont {A.}~\bibnamefont {Gelman}},\ and\ \bibinfo {author} {\bibfnamefont {W.~R.}\ \bibnamefont {Gilks}},\ }\bibfield  {title} {\bibinfo {title} {Efficient metropolis jumping rules},\ }\href@noop {} {\bibfield  {journal} {\bibinfo  {journal} {Bayesian Statistics}\ }\textbf {\bibinfo {volume} {5}},\ \bibinfo {pages} {599} (\bibinfo {year} {1996})}\BibitemShut {NoStop}%
\bibitem [{\citenamefont {Bouvier}\ \emph {et~al.}(2023)\citenamefont {Bouvier}, \citenamefont {Corenflos}, \citenamefont {Bortoli}, \citenamefont {Durmus} \emph {et~al.}}]{BlackJAX}%
  \BibitemOpen
  \bibfield  {author} {\bibinfo {author} {\bibfnamefont {L.}~\bibnamefont {Bouvier}}, \bibinfo {author} {\bibfnamefont {A.}~\bibnamefont {Corenflos}}, \bibinfo {author} {\bibfnamefont {V.~D.}\ \bibnamefont {Bortoli}}, \bibinfo {author} {\bibfnamefont {A.}~\bibnamefont {Durmus}}, \emph {et~al.},\ }\href@noop {} {\bibinfo {title} {Blackjax: Composable bayesian inference in jax}} (\bibinfo {year} {2023}),\ \Eprint {https://arxiv.org/abs/2310.07879} {arXiv:2310.07879 [stat.CO]} \BibitemShut {NoStop}%
\end{thebibliography}%

\end{document}